\documentclass[12pt,final]{elsarticle}

\makeatletter
\def\ps@pprintTitle{%
  \let\@oddhead\@empty
  \let\@evenhead\@empty
  \let\@oddfoot\@empty
  \let\@evenfoot\@empty}
\makeatother

\usepackage{amssymb}
\usepackage{eurosym}

\usepackage{graphicx} 
\usepackage{makeidx} 
\usepackage{amsfonts} 
\usepackage{latexsym} 
\usepackage{amsmath}  
\usepackage{amsthm}
\usepackage{color}    
\usepackage{hyperref} 
\usepackage{multicol} 
\usepackage{epstopdf}
\usepackage{subfigure}
 \usepackage{setspace} 
 \usepackage{float} 

\newcommand{\mm}[3]{\renewcommand{\arraystretch}{0.8}\begin{array}[t]{c}\mbox{#1} \\ #2\end{array}\begin{array}[t]{c}#3\end{array}
\renewcommand{\arraystretch}{1}}

\newcommand{\bfg}[1]{\mbox{\boldmath $#1$\unboldmath}}

\newcommand{\fraca}[2]{\displaystyle\frac{#1}{#2}}

\def \R {{\rm I\kern -2.2pt R\hskip 1pt}}

\journal{}

\begin{document}

\begin{frontmatter}



\title{Offshore Wind Farm Layout Optimization using Mathematical Programming Techniques}


\author[a]{Beatriz P\'erez}
\author[a]{Roberto M\'inguez}
\author[a]{Ra\'ul Guanche}

\date{2013}

\address[a]{Environmental Hydraulics Institute
$``$IH-Cantabria$"$}

\begin{abstract}
Offshore wind power is a renewable energy of growing relevance in
current electric energy systems, presenting favorable wind
conditions in comparison with the sites on land. However, the
higher energy yield has to compensate the increment in
installation and maintenance costs, thus the importance of
optimizing resources. One relevant aspect to increase
profitability is the wind farm layout. The aim of this paper is to
propose a new method to maximize the expected power production of
offshore wind farms by setting the appropriate
 layout, i.e. minimizing the wake effects. The method uses a sequential procedure for global optimization consisting of two steps:
 i) an heuristic method to set an initial random layout configuration, and ii) the use of nonlinear mathematical
 programming techniques for local optimization, which use the random layout as an initial solution. The method takes
 full advantage of the most up-to-date mathematical programming techniques while performing a global optimization approach,
 which can be easily parallelized. The performance of the proposed procedure is tested using the German
 offshore wind farm Alpha Ventus, located in the North Sea, yielding an increment of expected annual power production of 3.758$\%$ with respect to the
  actual configuration. According to current electricity prices in Germany, this constitutes an expected profit increment of
  almost 1M\euro{} per year.
\end{abstract}


\end{frontmatter}



\section{Introduction}
%
%
Wind energy is one of the most profitable renewable energy
sources, constituting a proven technology to meet current and
future electricity demands. Most of the operating wind farm
turbines are on land, however an important part of the future
expansion of wind energy, mainly in Europe, is expected to come
from offshore sites.

Offshore wind conditions are favorable with respect to sites on
land, presenting stronger and steadier wind speeds. However, the
advantages with respect to the potential wind resource contrast
with the increments of installation and maintenance costs, which
must be somehow compensated. This reason has motivated scientist
and engineers to focus on optimizing offshore wind farm project
designs, focusing on different aspects such as location
\cite{MinguezMML:11}, installation, layout
\cite{Mosetti:94,Ozturk:03,Grady:04,Elkinton:06}, availability,
operation and maintenance \cite{RademakersBZB:03,MinguezMCG:11},
etc... Note that although all these aspects are relevant, in this
study we focus only on the layout optimization.

Once the wind off-shore resource is probabilistically
characterized at a particular location, it is possible to
strategically position the turbines in order to minimize expected
wake effect losses, thus maximizing the expected efficient energy
production. This problem is referred to as optimizing the layout
of a wind farm. Note that when the wind goes though any turbine, a
wake effect is induced downstream decreasing wind speed and
increasing wind turbulence. This produces a reduction of energy
production in all turbine located within the area of influence of
the wake.

Different studies on layout optimization have been proposed in the
literature. The first work that addresses this problem is
\cite{Mosetti:94}, which use genetic algorithms to determine the
positions of wind turbines that provide the maximum energy
extraction with the minimum installation costs. A decade later,
\cite{Ozturk:03} propose the use of an heuristic methodology based
on
 Greedy in order to maximize profits
rather than the energy produced in the wind farm.
\cite{Donovan:06} formulates the generalized vertex parking
problem (GVP) and
 obtain the maximum energy production subject to several constraints. However, the author does not
 clearly state which wake model is used for the study.
\cite{Kusiak:10} proposes a multi-objective optimization problem
using  genetic algorithms, maximizing the energy production and
minimizing the failure of the limitations.
\cite{Elkinton:06} develops within the auspicious of the Offshore
Wind Farm Optimization (OWFLO) project, a more comprehensive study
combining an energy production model (taking into account wake
effects, electrical losses and turbine availability) with offshore
wind farm component cost models. This project aims to pinpoint the
major economic hurdles present for offshore wind farm developers
by creating an analysis tool that unifies offshore turbine
micrositing criteria with efficient optimization algorithms. Finally,
\cite{Chowdhury:12} proposes the Unrestricted Wind Farm Layout Optimization (UWFLO) methodology,
that addresses critical aspects of optimal wind farm planning. It simultaneously
determines the optimum farm layout and the appropriate selection of turbines
(in terms of their rotor diameters) that maximizes the net power generation.

To our knowledge, all optimization algorithms proposed for layout
optimization are based on heuristic procedures, specially Genetic
Algorithms \cite{Elkinton:08}. \cite{SerranoGCRB:10} presents
an evolutive algorithm to
optimize the wind farm layout onshore. The algorithm's optimization process is based on a global
wind farm cost model using the initial investment and the present value of the yearly
 net cash flow during the entire wind-farm life span. \cite{SaavedraMorenoSPP:11}
 proposes a novel evolutionary algorithm for optimal positioning of wind turbines in wind
 farms. For this case, a realistic model for the wind farm is considered, which includes
orography, shape of the wind farm, simulation of the wind speed and direction, and costs of installation,
connection and road construction among wind turbines.
\cite{ErogluU:12} introduces an ant colony algorithm for maximizing the expected energy output.

The main idea of these methods is to generate,
evaluate, and select possible solutions based on different
principles, depending on the type of method, until the algorithm
is unable to find a better solution. Basically, these methods
focus on finding an acceptable solution in an attempt to capture
the global optimum. However, they use simplifying assumptions and
do not ensure neither local nor the global optimum, which means
that most of the times the solutions obtained do not even hold the
Karush-Kuhn-Tucker optimality conditions (see
\cite{Vanderplaats:84,BazaraaSS:93}).
In particular, and regarding the layout optimization problem, existing approaches
discretize the possible locations of turbines over a predefined grid which limits the feasible region of possible locations
considerably.

The selection of heuristic instead of mathematical programming techniques for layout optimization
has been based on two main assumptions:
\begin{enumerate}
  \item The computational time of gradient-based mathematical programming methods is prohibitive to
  solve these kinds of problems.

  \item The optimal location problem is non-convex, and gradient-based methods provide local solutions. Thus, depending
  on the initial solution used to start running these algorithms, the global optimum may be skipped.
\end{enumerate}

The aim of this paper is to drop these assumptions by presenting a combined method, heuristic versus gradient-based,
to obtain the best offshore wind farm layout over a pre-specified area. The proposed procedure
takes full advantage of the state-of-the-art nonlinear programming solvers. Since the global optimum must
 lie in a convex subregion, which may be identified by the mathematical programming solvers,
 we look for the global optimum by restarting heuristically the initial solution used to run gradient-based
  solvers. The proposed methodology has the following advantages:
  \begin{enumerate}
    \item Current state-of-the-art nonlinear mathematical programming solvers are more reliable, numerically robust,
    and computationally efficient.

    \item Nonlinear mathematical programming solvers allow including alternative constraints easily, or objective functions,
    which do not alter the flow of the methodology.

    \item The heuristic method used to generate initial solutions is capable of searching convex
     subregions. This allows tackling
    non-convexities.

    \item It is easy to include parallelization features in order to increase computational efficiency and reduce
    computational costs.

    \item The final solution holds the Karush-Kuhn-Tucker optimality conditions.

\item It does not require reducing the feasible solution region by gridding the possible location area.

  \end{enumerate}

The rest of the paper is structured as follows. Section~\ref{wakemodel} justifies the wake model selection.
Section~\ref{optlay} and Section~\ref{optimization_cap} present
the layout optimization methodology formulating
 the mathematical statement of the
problem and the solution algorithm. In Section~\ref{Aventus}, the
proposed method is applied using the German offshore wind farm,
Alpha Ventus, and finally, in Section~\ref{concl} some relevant
conclusions
 are duly drawn.

\section{Wake models}\label{wakemodel}

A wake is the downstream region of disturbed flow, usually
turbulent, caused by a body moving through a fluid. In the case of
wind turbines, the wind forces the blades to rotate, thus
generating the mechanical energy which is subsequently converted
to electricity. This energy extraction decreases the wind speed
and increases turbulence at the rear of the turbine, which reduces
the energy production at downwind turbines.



Several studies which carry out extensive comparisons between
different wake models (see
\cite{Vanluvanee:06,Shorensen:08,WebWinpro,Renkema:07}) allow
concluding that there is a high uncertainty in all models
performance. However, \cite{Vanluvanee:06}, based on the findings
from his work, recommends the N.O. Jensen model be used for the
energy predictions in offshore wind farms, as it offers the best
balance between positive and negative prediction errors. This is
the model selected for this study.


For a location $i$, located on the downstream wake induced by
turbine $j$, and at a distance $d_{ij}$ projected on the wind
direction between turbine $j$ and the point of study $i$, the wake
velocity deficit $D_{v_{ij}}$ is given by the following
expression:
\begin{equation}\label{def_vel}
D_{v_{ij}}=1-\frac{v_{i}}{v_j}=\frac{(1+\sqrt{1-C_{t_j}})}{\bigl(1+\frac{k\cdot
d_{ij}}{R}\bigl)^2},
\end{equation}
where $v_i$ is the velocity at location $i$ within the wake, $v_j$
the wind speed reaching turbine $j$, $C_{t_j}$ is the thrust
coefficient associated with velocity $v_j$, $k$ is the decay
factor, and $R$ is the rotor radius.

The decay factor $k$ describes how the wake breaks down by
specifying the growth of the wake width per meter traveled
downstream. The determination of $k$ is sensitive to factors
including ambient turbulence, turbine induced turbulence and
atmospheric stability. In a simplified manner, the calculation is
performed through the following equation:
\begin{equation}\label{eq_k}
k=\frac{A}{\ln{z\overwithdelims()z_0}}=\frac{1}{2\ln{h\overwithdelims()z_0}},
\end{equation}
where $z$ is the height of the turbine, $A$ a constant
approximately equal to 0.5, and $z_0$ is the surface roughness.
Parameter $z_0$ is crucial in the decay coefficient. There are
numerous recommendations \cite{Vanluvanee:06,Wasp:11,Thogersen:11}
for its selection. Typical values for different kinds of terrains
are given \cite{Wasp:11}.



This model was first described by \cite{Jensen:84} and further
developed by \cite{Katic:86}. It is used in several commercial
softwares, such as, WAsP~\cite{Wasp:11}, Garrad Hassan
WindFarmer~\cite{GH:04}, and WindPRO~\cite{WebWinpro}.

\section{Layout problem definition}\label{optlay}
This study aims to determine the optimal layout of the wind
turbines inside an offshore wind farm in order to reduce the wake
effects as much as possible. Since we propose to face this problem
using mathematical programming techniques, we start by defining
the four basic elements required to state any optimization problem
\cite{CastilloCPGA:01,ConejoCMG:06}: i) data, ii) problem
variables, iii) constraints, and iv) the objective function.

Once the main elements of the problem are described, we explain in
detail the combined heuristic versus mathematical programming
strategy used to solve it.

\subsection{Data}
The data constitutes the information which is known and required
to set and appropriately calculate the objective function and
constraints. For this particular case, it can be classified in the
following sets:
\paragraph{\bf Wind data} This set includes all wind-related parameters associated with the
location:
  \begin{enumerate}
    \item Wind data at 10 meters height in the study area ($v_{10}$), including both wind speed and directional information.
     This data is critical to correctly predict on the energy production and evaluate
     wake losses. It can be based on i) instrumental measurements in the field, which usually provide
      accurate information although of short length, ii) reanalysis data, which constitutes an alternative
      to providing long records \cite{MenendezTCGFFML:11}, or iii) a combination of both, i.e. reanalysis data
       calibrated using instrumental information from satellite or from the field \cite{MinguezMML:11}.
    \item Coefficient of roughness length or surface roughness ($z_0$), which is required to evaluate the wind speed at different
    heights.
  \end{enumerate}

\paragraph{\bf Turbine data} This set includes all parameters related to the specific turbine:
\begin{enumerate}
  \item Hub height ($z$). This information allows us to calculate the required wind speed at the height using an appropriate wind profile.

  \item Rotor diameter or radius ($D,\;R$). The energy produced by the turbine is dependent on this value, and also affects the form of the wake.

    \item Thrust coefficient ($C_t$). This information is turbine specific, and is usually given as a curve,
    which depends on the wind speed at the hub $v$.

    \item Power curve ($P_w$). This curve defines the energy produced by the turbine as a function of the
the wind speed at the hub $v$. It includes information on the
 control mechanisms.

\end{enumerate}

\paragraph{\bf Wake effect data} This set includes all parameters required by the wake effect model
not previously mentioned:
\begin{enumerate}
  \item The decay factor $k$.
  \item The minimum distance where the wake model is considered to work appropriately ($d_{\rm min}$).
\end{enumerate}

\paragraph{\bf Wind farm data} This set includes all parameters associated with the wind data:
\begin{enumerate}
  \item Area where in which turbines can be located.
  \item Number of turbines ($N_T$) to be allocated within the wind farm area.
\end{enumerate}

\subsection{Problem variables}
The variables constitute the decisions to be made, which for this
particular case are the exact location coordinates of each turbine
$(x_i,y_i);\;\forall i=1,\ldots,N_T$. These variables are
integrated into the variable decision vector ${\bfg
x}\in\Re_{N_t\times2}$ as follows:
\begin{equation}\label{variables_opt}
{\bfg x}=\left[
\begin{array}{cc}
  x_1 & y_1 \\
  x_2 & y_2 \\
  \vdots & \vdots \\
  x_{N_T} & y_{N_T}
\end{array}
\right].
\end{equation}

Note that besides wake effects, the appropriate layout of any wind
farm is influenced by additional factors, such as, water depth
(foundation costs). Water depth could be included into the
decision variable vector as an additional coordinate
($z_i;\;\forall i=1,\ldots,N_T$). Its consideration would require
to define the bathymetry over the wind farm area as data.
Nevertheless, this is out of the scope of the paper, and
constitutes a subject for further research

\subsection{Constraints}
The set of constraints determine which decisions are admissible,
i.e. define the feasible region of the problem variables. In this
paper we have considered two types of constraints:
\paragraph{\bf Minimum distance}
The wake model selected (\cite{Jensen:84}) is known to provide
appropriate results for distances higher than four rotor
diameters, and for safety reasons, the minimum distance between
turbines within the wind farm is limited to four rotor diameters
($d_{min}=4D$). This restriction can be mathematically expressed
as follows:
\begin{equation}\label{restric1_opt}
\fraca{(x_i-x_j)^2+(y_i-y_j)^2}{(4D)^2}\geq 1;\;\forall i=1,\dots,(N_T-1);\:\forall j>i.
\end{equation}

Note that the manner in which constraint (\ref{restric1_opt}) is
defined facilitates gradient-based methods to converge because of
an adequate scaling, i.e. the variable units do not affect the
solution.

\paragraph{\bf Wind farm area limits}
Turbines must be allocated to a predefined area given as data. The
optimization process must ensure
 that turbines are within the required area. In this paper, we define this limiting area using 4 nodes (quadrilateral)
  Given these four coordinates $(x_i^L,y_i^L);\;i=1,2,3,4$
  as data, the constraint that ensures turbines remain inside this quadrilateral region is as follows:
%
%
\begin{equation}\label{determinante_opt}
\left|
\begin{array}{ccc}
x_i & y_i &1\\
x_{j}^L & y_{j}^L &1\\
x_{k}^L & y_{k}^L &1
\end{array}
\right| \geq0;\;\left\{
\begin{array}{l}
\forall i;\; j=1;\; k=2\\
\forall i;\; j=2;\; k=3\\
\forall i;\; j=3;\; k=4\\
\forall i;\; j=4; \;k=1.
\end{array}\right.
\end{equation}

Constraint~(\ref{determinante_opt}) requires that the area of the
four triangles formed by each node-turbine with each pair of
consecutive boundary nodes (measured in counterclockwise
direction) is always greater than zero.

It was mentioned before that besides wake effects, the appropriate
layout of any wind farm is influenced by additional factors. Some
of these additional aspects are visual impact, environmental
factors, tourism and legal approval, among many others. We assume
that the selection of the location limiting area takes into
account all these factors.

 Note that both restrictions (\ref{restric1_opt}) and (\ref{determinante_opt}) are linear inequality
constraints.

\subsection{Objective function: Annual Energy Production (AEP)}
Finally, a function that allows us to characterize how good or bad
any decision is must be defined. For this particular case, we have
decided to use the expected Annual Energy Production (AEP), which
takes into consideration the wake effects and allows quantifying
the benefits in terms of profit from the positioning strategy.

The wake effect depends on wind speed magnitudes and directions.  For this reason the calculation
of the expected AEP requires the definition of the wind speed probability density function
conditional on the direction. Assuming that the conditional distribution is Weibull, which
is widely accepted for wind data, the probability density function is defined as follows:
\begin{equation}\label{weibull_1}
f_{V|\theta}(v)=\frac{\delta(\theta)}{\lambda(\theta)}{v\overwithdelims()\lambda(\theta)}^{\delta(\theta)-1}e^{-{v\overwithdelims()\lambda(\theta)}^{\delta(\theta)}},
\end{equation}
where $v$ represents the wind speed at the hub height, $\delta$ and $\lambda$ are the shape and
 scale parameters of the Weibull distribution, respectively, and $\theta$ is the wind direction.

The parameters of the distribution (\ref{weibull_1}) need to be estimated from real data, which in our case
 consist of hourly time series of wind speeds and directions. To facilitate calculations and speed up the objective
function evaluation for a given layout configuration, we divide
the possible directions into a number of sectors $n_s$. Data from
each sector is fitted to a Weibull distribution. The selection of
the number of sectors is very important as a small number is
favorable for the Weibull fitting but the errors from the wake
effect increase because the mean direction is used as the
representative for each sector. On the other hand, increasing the
number of sectors decreases the quality fit, however it also
decreases the wake effect errors. Numerical tests performed with
different number of sectors \cite{Perez:12} allows us to conclude
that the selection of 12 sectors of $30^\circ$ width provides the
best compromise in terms of i) Weibull fitting quality, ii) wake
effect errors, and iii) computational time, bounding the errors
with respect to the expected AEP below $0.5$\%. This result is in
accordance with conclusions given in \cite{Vanluvanee:06}.

\begin{figure}[!h]
\begin{center}
\includegraphics*[width=0.7\textwidth]{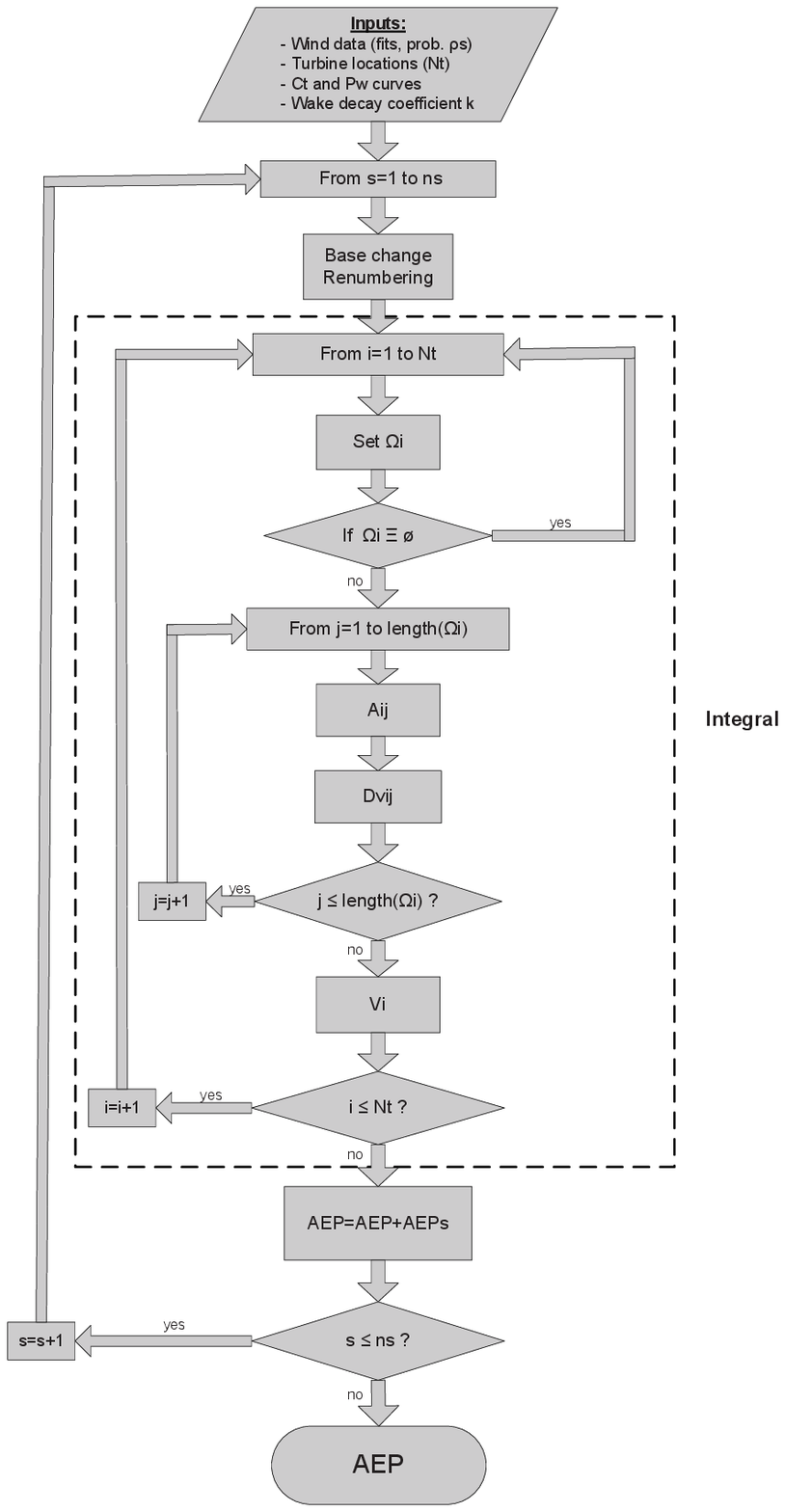}
\caption{\label{diagrama_flujo1} Flow diagram: calculating the
Annual Energy Production (AEP).}
\end{center}
\end{figure}

Once the number of sectors $n_s$ is defined, the expected Annual
Energy Production (AEP) is calculated as follows:
\begin{equation}\label{AEP}
\mbox{AEP}=8760\sum_{s=1}^{n_s}\rho_s\sum_{i=1}^{N_T}\left[\int_{v_{c_i}}^{v_{c_o}}{f_{V|\theta_s}(v)P_w(v_i)}\,dv\right],
\end{equation}
where 8760 is the mean number of hours per year, $\rho_s$ is the probability of the wind to
be within sector $s$,
$v_{c_i}$ and $v_{c_o}$ are, respectively, the cut-in and cut-out velocities defined
 in the turbine power curve, and $P_w(v_i)$ is the power produced by turbine $i$ for its
corresponding wind speed.
 Note that $v_i$ comes from the perturbed wind speed $v$ at the hub height decreased by the wake effect.
 These effects depend on the wind direction and the positioning of each turbine compared to the
 others. Note that the objective function calculation is based on the sum of the energy yield by
each turbine for each sector, as shown in equation (\ref{AEP}).

The calculation of the expected AEP according to (\ref{AEP}) is a
key step for the correct performance of the proposed optimization
method. To facilitate the understanding  of the process, its flow
diagram is shown in Figure~\ref{diagrama_flujo1}.

In addition, the most relevant steps within the algorithm are described below for a given
directional sector $s$:

\paragraph{\bf Wind farm reorientation and renumbering}
This step consists of a base change to rotate the cartesian axis
by an angle $\theta$ measured from the north to the average
direction of the study. The
rotation can be expressed mathematically as:
\begin{equation}\label{cambio_base1}
{\bfg x}^\prime =
\left(\left[
\begin {array}{cc}
\cos\theta_s & -\sin\theta_s\\
\sin\theta_s & \cos\theta_s
\end {array}
\right] {\bfg x}^T\right)^T.
\end{equation}

\begin{figure}[!h]
\begin{center}
\includegraphics*[width=0.62\textwidth]{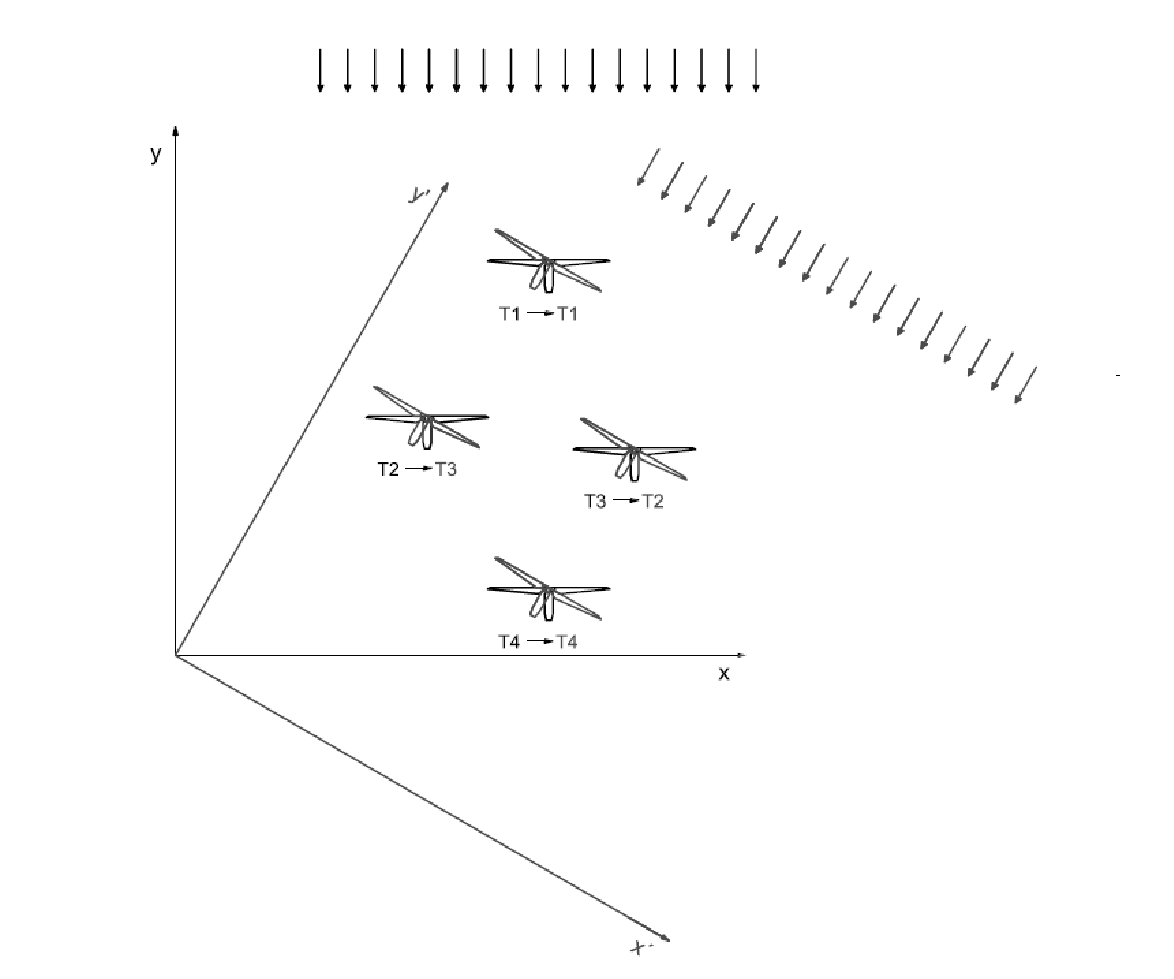}
\caption{\label{reordenacion} Renumbering of the turbines after the base change.}
\end{center}
\end{figure}

Once the new coordinates are calculated, it is very simple to renumber turbines
ranked according to their
vertical coordinate $y^\prime$ in descending order (see Figure~\ref{reordenacion}).

Note that this step allows us to rapidly select the possible
turbine set $\Omega_i$ affecting each turbine $i$ due to wake
effect, that is, those with cardinality lower than $i$, i.e.
$\Omega_i=\{1,2,\ldots,i-1\}$.

\paragraph{\bf Velocity evaluation due to the wake effect}
For all turbines, it is required the true velocity at the rotor
must be calculated, which may decrease due to the wake effect.
Given a specific unperturbed velocity $v$ and turbine $i$, this
process entails the following steps which must be repeated with
all turbines belonging to set $\Omega_i$:
\begin{itemize}
  \item {\bf Step 1: Cross sectional area intersection.} Before
  evaluating the wind speed deficit, one must calculate the area of
the wake produced by turbine $j$
  that intersects with the rotor swept area of the downstream turbine $i$, i.e. $A_{ij}$.
  This is a geometric problem in which there are four possible cases as shown in
  Figure~\ref{casos}.
%
%
\begin{figure}[!h]
\begin{center}
\includegraphics*[width=0.60\textwidth]{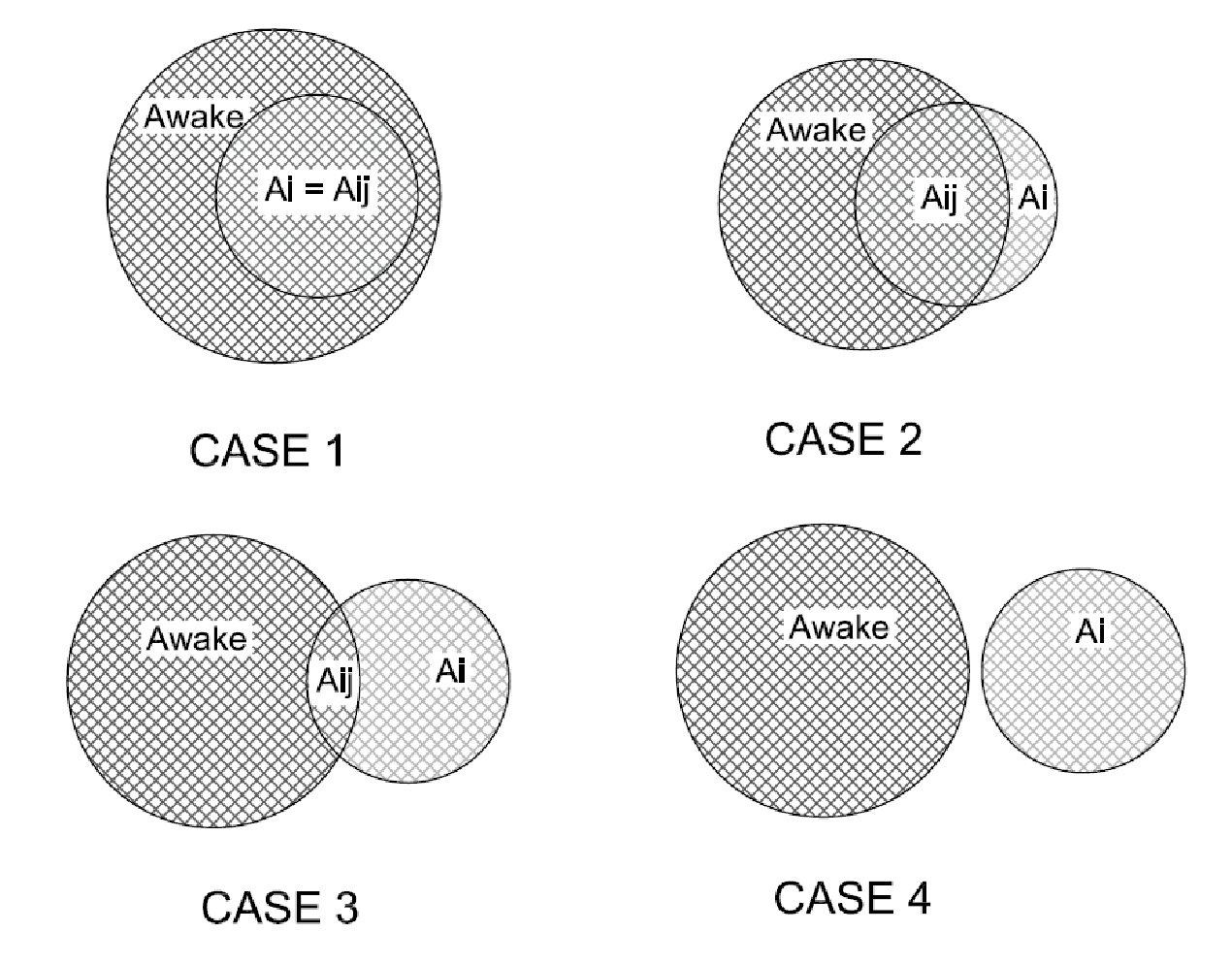}
\caption{\label{casos} Different possibilities for cross sectional
intersection problem area.}
\end{center}
\end{figure}

\item {\bf Step 2: Deficit velocity evaluation.} Once the intersection area $A_{ij}$ is available, the deficit
of wind velocity due to the wake effect is calculated according to
\cite{Wasp:11} as follows:
\begin{equation}\label{def_vel_2}
D_{v_{ij}}=\frac{(1+\sqrt{1-C_{t_j}})}{\bigl(1+\frac{k\cdot
d_{ij}}{R}\bigl)^2}\cdot\frac{A_{ij}}{A_i},
\end{equation}
where $A_i$ is the rotor area, and the thrust coefficient
$C_{t_j}$ is calculated using the curve for the velocity at
turbine $j$, which was previously calculated. Thus the importance
in calculating the velocities from upstream to downstream. Note
that (\ref{def_vel_2}) is a slight modification of expression
(\ref{def_vel}) that takes into account the extent
 to which the turbine $i$ is affected by the wake of the turbine $j$.

At this step of the process, we have calculated all deficits $D_{ij};\;\forall i\in \Omega_i$ affecting
turbine $i$ velocity. \cite{Katic:86} proposed an energy balance to calculate the cumulative
deficit produced by several turbine wakes, so that the total speed
deficit for turbine $i$ is computed as follows:
\begin{equation}\label{combinacion_est}
(D_{v_{i}})^2=\sum_{\forall j\in \Omega_i}(D_{v_{ij}})^2.
\end{equation}

According to (\ref{def_vel}), the unperturbed wind speed at the
hub height $v$, affecting the wind farm, due to the wind farm wake
effect for turbine $i$ turns into:
\begin{equation}\label{veli}
    v_i = v(1-D_{v_{i}}).
\end{equation}

\end{itemize}

\paragraph{\bf Power output evaluation}
The power produced by turbine $i$ is obtained from the given power curve as a function
of the calculated wind speed $v_i$.

Note that the previous steps were explained only for a given i)
directional sector, ii) turbine $i$,
 and iii) unperturbed velocity $v$. However,
this process is embedded in an integral over the range of possible
wind speeds, which is numerically evaluated using the trapezoidal
quadrature
 formula. Numerical tests performed using different quadrature (Simpson) methods and size steps
  show that the trapezoidal rule using a step size $\Delta v=0.1$ m$/$s provides the best counter-balance
  between small errors (below $0.01$\%) and computational efficiency (see \cite{Perez:12}).

From a mathematical point of view, it would be very easy to
include additional factors on the objective function to account
for the water depth (foundation costs) or electrical connection to
the grid (cable costs). However, this is out of the scope of the
paper.


\section{Layout optimization}\label{optimization_cap}
Once the four elements of the optimization problem are defined.
The mathematical programming definition is as follows:
\begin{equation}\label{d}
   \mm{Maximize}{{\bfg x}}{AEP,}
\end{equation}
given by (\ref{AEP}) and subject to constraints
(\ref{restric1_opt}) and (\ref{determinante_opt}).

According to the type of objective function and constraints, this
problem is a nonlinear mathematical programming problem with
linear inequality constraints. It can be efficiently solved using
any of the available solvers for nonlinear programming subject to
constraints, for instance, solver MINOS \citep{MurtaghS:98} or
CONOPT \citep{Drud:96} under GAMS \citep{BrookeKMR:98}, the Trust
Region Reflective Algorithm under Matlab
\citep{ColemanL:94,ColemanL:96}, also capable of dealing with
nonlinear equality constraints and upper and lower bounds through
the function \verb"fmincon", or interior-point barrier and active
set techniques implemented on function \verb"ktrlink"
\cite{ByrdNW:06}, where each algorithm addresses the full range of
nonlinear optimization problems, and each is constructed for
maximal large-scale efficiency.

Numerical tests performed using functions \verb"fmincon" and
\verb"ktrlink" show that these algorithms are capable of finding
solutions holding the KKT optimality conditions in a reasonable
computational time. However, the problem is that they are suited
for finding local minima, and depending on the initial layout
configuration used to run the algorithms, different local minima
could be found. To overcome this difficulty, we propose a combined
method based on two basic steps:
\begin{description}
  \item[Heuristic initial solution:] In an attempt to identify all possible convex subregions, an initial
stochastic solution ${\bfg x}^s$ holding constraints
(\ref{restric1_opt}) and (\ref{determinante_opt}) is
 generated.

  \item[Local minima search:] Run any mathematical programming technique using the sample solution ${\bfg x}^s$ as a starting value.
\end{description}

Repeating this process allows us to explore all possible local
solutions. Although
 a finite number of iterations does not guarantee that the global optimum is achieved,
the probability of succeeding increases with the number of times the proposed sequential procedure is
 performed. The main idea is to make as many repetitions as possible depending on different factors, such as
 the time available to carry out the analysis, computational resources, number of turbines, ... Nevertheless,
we will always obtain a good solution, and several rules of thumb
may be used to decide reasonable stopping criteria. For example,
we could stop the process if we do not improve the best solution
obtained so far during a pre-specified number of iterations.

Note that the structure of the problem would allow us to
parallelize the process easily, each core may repeat this
heuristic-local search process independently. The only requirement
is to share the best solution found between cores.

\subsection{Heuristic initial solution}
The first step of the proposed procedure consists of the
simulation of an initial solution, which is used as starting point
to run gradient-base algorithms. The aim of this stage is to cover
all possible local solutions and facilitate gradient-based
algorithm performances. For these reasons, the heuristic
generation must fulfill the following conditions:
\begin{enumerate}
  \item The initial solution must cover all possibilities, thus the initial location must be random, being able to
locate turbines anywhere within the wind farm area.
  \item The initial solution must hold the linear inequality constraints (\ref{restric1_opt}) and (\ref{determinante_opt}).
  \item If there were no constraints on the wind farm location area, the best known solution to avoid wake
 effects to locate turbines as far away from each other as possible. Thus, we try to cover all study area as much as possible.
\end{enumerate}

In order to fulfill these requirements, we use a two step
procedure. The method aims to maximize the use of the area
assigned by random spread of the turbines, taking into account
constraints (\ref{restric1_opt}) and (\ref{determinante_opt}):
\begin{description}
  \item[Uniformly random simulation] Simulate $N_T$ uniformly distributed random numbers
$\xi_i\sim U(-1,1)$ and $\eta_i\sim U(-1,1);\; \forall
i=1,\ldots,N_T$. Note that the
probability of locating the coordinate $(\xi_i,\eta_i)$ in any
point within the rectangle is the same.
%

  \item[Rectangle transformation] Transform those points from the regular rectangle into the
wind farm area defined by points  $(x_i^L,y_i^L);$ $i=1,2,3,4$ as follows:
\begin{gather}\label{simul5}
\begin{split}
x_{i,0}^s=\sum_{l=1}^{4}N_l(\xi_i,\eta_i)x_l^L;\;\forall i=1,\ldots,N_T\\
y_{i,0}^s=\sum_{l=1}^{4}N_l(\xi_i,\eta_i)x_l^L;\;\forall i=1,\ldots,N_T,
\end{split}
\end{gather}
where functions $N_l;\;\forall l=1,2,3,4$ are equal to:
\begin{eqnarray}
N_1(\xi,\eta)=\frac{1}{4}(1-\xi)(1-\eta)\\
N_2(\xi,\eta)=\frac{1}{4}(1+\xi)(1-\eta)\\
N_3(\xi,\eta)=\frac{1}{4}(1+\xi)(1+\eta)\\
N_4(\xi,\eta)=\frac{1}{4}(1-\xi)(1+\eta).
\end{eqnarray}


 \item[Turbine widespread] In order to fully exploit the wind farm area, we try to spread
turbines out all over the limiting area. Thus, we perform a
Delaunay triangulation \cite{BarberDH:96} using the simulated
points $(x_{i,0}^s,y_{i,0}^s);\;\forall i=1,\ldots,N_T$, and then
maximize the sum of triangle areas solving the following
optimization problem:
\begin{equation}
\mm{maximize}{(x_i^s,y_i^s);\;\forall i=1,\ldots,N_T}{\displaystyle \sum_{\forall t}A_t}
\end{equation}
subject to (\ref{restric1_opt}), (\ref{determinante_opt}) and
\begin{equation}\label{cnocross}
    A_t>0\; ; \forall t,
\end{equation}
using as starting variables $(x_{i,0}^s,y_{i,0}^s);\;\forall
i=1,\ldots,N_T$. $A_t$ is the area of triangle $t$, which can be
calculated using the determinant formula:
\begin{equation}\label{fobjareas}
A_t = \fraca{1}{2}\left|
\begin{array}{ccc}
x_{t_1} & y_{t_1} &1\\
x_{t_2} & y_{t_2} &1\\
x_{t_3} & y_{t_3} &1
\end{array}
\right|\; ; \forall t.
\end{equation}

Constraint (\ref{cnocross}) avoids triangle overlapping and edge
crossing when they are distorted by moving their vertices. The
optimal solution $(x_{i}^s,y_{i}^s);\;\forall i=1,\ldots,N_T$ of
this problem hold constraints (\ref{restric1_opt}) and
(\ref{determinante_opt}), and is known in advance, since the
maximum sum of areas must be equal to the wind farm limiting area,
as shown in Figure~\ref{delaunay}. This problem is a nonlinear
mathematical programming problem easily solvable using any of the
previously mentioned algorithms.

\begin{figure}[!h]
\begin{center}
\includegraphics*[width=0.9\textwidth]{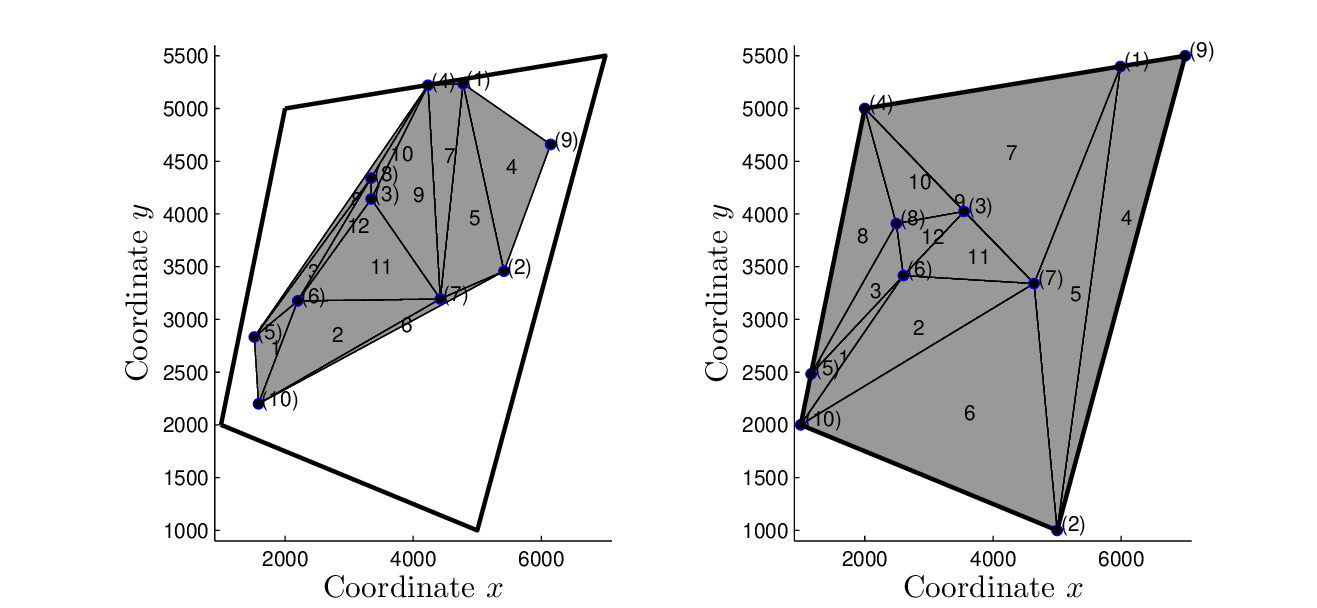}
\caption{\label{delaunay} Initial and optimal Delaunay
triangulations.}
\end{center}
\end{figure}

\end{description}

This solution ${\bfg x}^s=(x_{i}^s,y_{i}^s);\;\forall
i=1,\ldots,N_T$ holds the three required conditions and is used as
the starting point to run the gradient-base algorithm.

\subsection{Local minima search}
Once the initial solution from the heuristic procedure is
obtained, the nonlinear mathematical programming algorithm is
executed.

Note that from a practical point of view, and according to results
obtained from numerical tests using different algorithms, we
decided to slightly change the local maxima search strategy. Thus,
we use function \verb"ktrlink" within Matlab, repeating the
combined process a fixed number of times $M$. Once the best
solution from the $M$ iterations is achieved, this is the
candidate to be the global solution. Then we run solver
\verb"fmincon" to slightly improve results.

%
\begin{figure}[!h]
\begin{center}
\includegraphics*[width=0.9\textwidth]{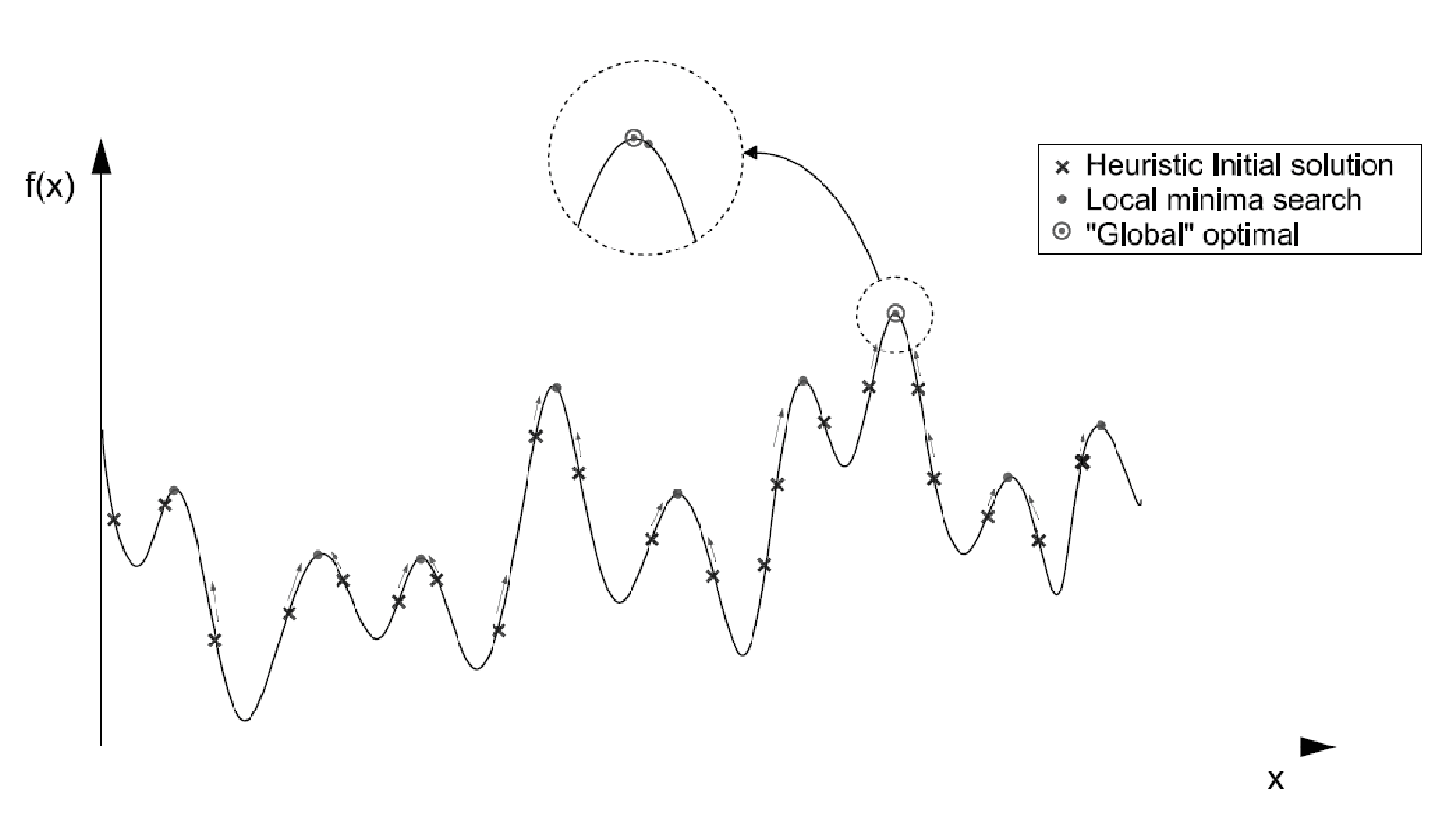}
\caption{\label{explica} Graphical interpretation of the combined
heuristic-gradient base layout optimization strategy.}
\end{center}
\end{figure}

The reason for this selection is graphically explained in
Figure~\ref{explica}. We initially use function \verb"ktrlink"
because it obtains a local maximum faster than \verb"fmincon".
However, once the ``global'' optimal ${\bfg x}_0^\ast$ is
achieved, function \verb"fmincon" slightly improves
 the objective function providing the final ``global '' optimal solution ${\bfg x}^\ast$. Figure~\ref{explica}
 showns i) the starting points used to look for convex subregions,
ii) the local maxima and iii) the global maximum.

Note that although there is no guarantee that ${\bfg x}^\ast$ is
the true global solution, the chances of finding it increase as
$M$ increases. However this methodology, unlike others, guarantees
the optimality conditions of the solution. Furthermore, this
gradient-base process could be used to refine solutions obtained
from other heuristic approaches.

\section{Case study: Alpha Ventus}\label{Aventus}
To show the functioning and the potential of the proposed
methodology, we select as a case study the Alpha Ventus wind farm.
It was the first offshore wind farm to be constructed in open sea
conditions (North Sea), 60 kilometres away from the coast, in the
midst of extreme winds, weather and tides.

Technically, Alpha Ventus \cite{webAlphaVentus:11} is equipped with the most advanced
technologies, specifically designed for offshore wind farms. The
wind turbines are placed in a grid-like formation with gaps of
approximately 800 meters between each turbine, in a rectangle with
a total surface area of four square kilometres.
The wind farm has
two types of turbines: the Multibrid M5000 and the REpower 5M, two
of the largest models in the world.

\begin{figure}[!h]
\begin{center}
\includegraphics*[scale=0.67]{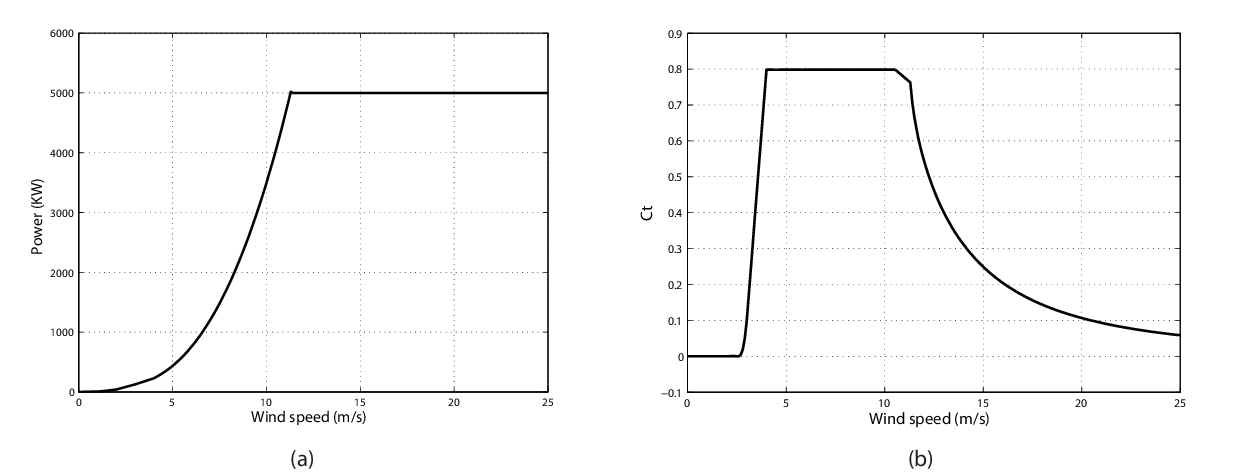}
\caption{\label{curvas} Characteristic curves of 5MW NREL Turbine:
a) Power output, and b) Trust coefficient.}
\end{center}
\end{figure}

For this application, it is assumed that the wind farm has 12
identical turbines, NREL 5 MW type~\cite{NREL}. The hub height and
rotor diameter are $z=90$ and $D=126$ meters, respectively. Power
curve ($v,P_w$) is illustrated in Figure~\ref{curvas} (a). It
 is a sigmoid function including three different
regions: i) $v<v_{c_i}=3$ m$/$s, ii) $v_{c_i}\le v<v_{r}=11.3$ m$/$s and $v_{r}\le
v<v_{c_o}=25$ m$/$s.
The thrust
curve ($v,\;C_t$) is shown in
 Figure~\ref{curvas} (b). Both curves must be provided by the manufacturer.

\begin{figure}[!h]
\begin{center}
\includegraphics*[width=0.5\textwidth]{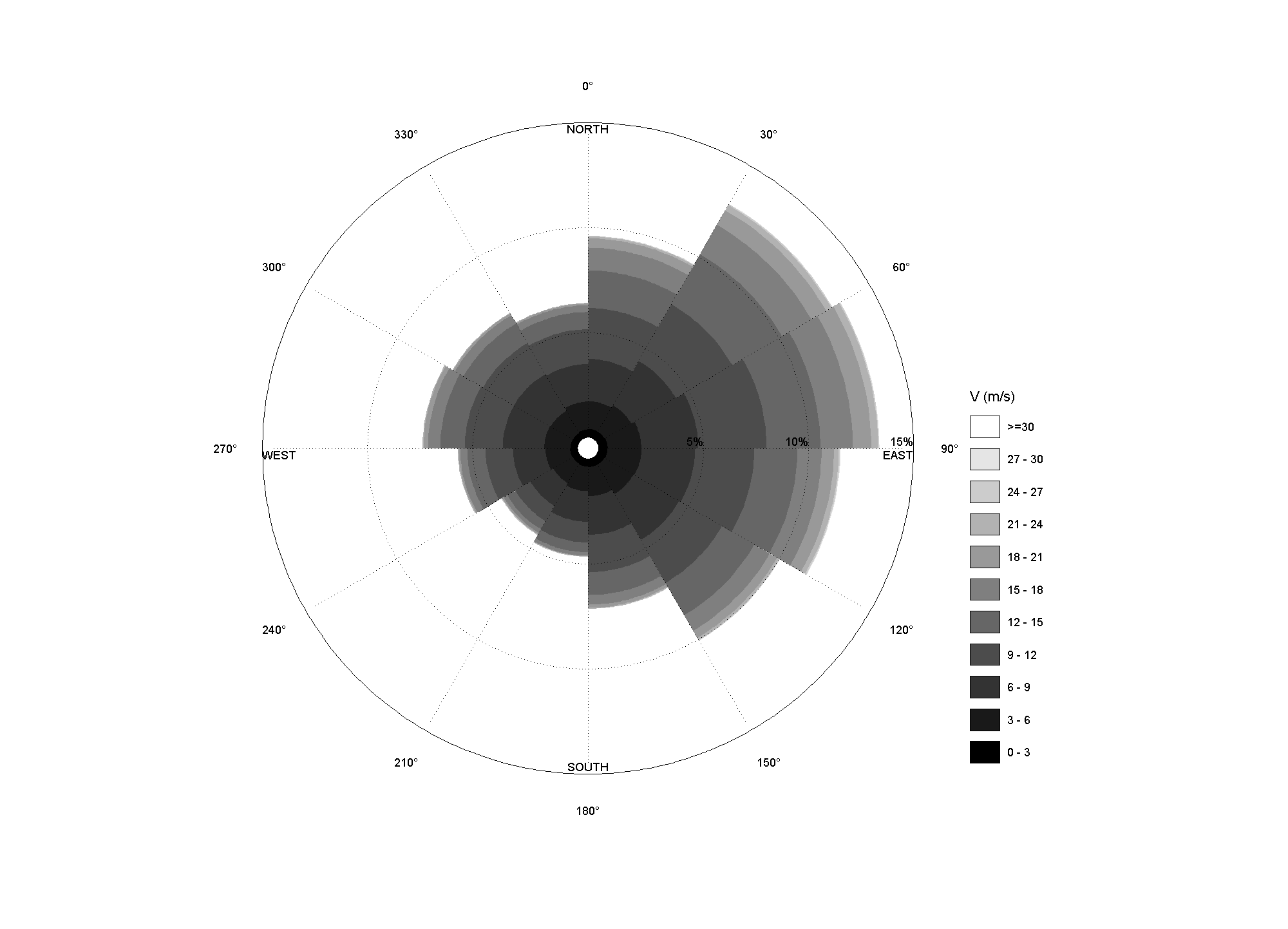}
\caption{\label{rosa} Wind rose for the Alpha Ventus location.}
\end{center}
\end{figure}

Wind data at 10 meters height in the study area is obtained from
the SeaWind database~\cite{MenendezTCGFFML:11}, which constitutes
an hourly wind reanalysis over a 15 km spatial resolution grid for
the entire 1989-2009 period, covering the South Atlantic European
region and the Mediterranean basin. The logarithmic wind speed
profile is chosen to achieve the wind speed at the hub height
$z=90$ meters. The equation of the profile is as follows:
\begin{equation}\label{logaritmica}
v=v_{10} {{\ln(Z/z_0)}\overwithdelims(){\ln(Z_{10}/z_0)}}
\end{equation}
where $z_0$ is the coefficient of roughness length, which has a
value for offshore zones of 0.0002 meters; $v$ and $v_{10}$ are
the wind speeds at 90 and 10 meters; $Z$ and $Z_{10}$ are the
heights at 90 and 10 meters. Figure \ref{rosa} shows the wind rose
in the area.

According to the available data, the expected annual AEP
considering no wake effect and total availability is equal to $306.9$~GWh.
Using (\ref{AEP}) and considering the wake effect, we obtain the results provided in Table~\ref{Alphaventuslayout} for
 the actual layout of Alpha Ventus.
 Results are given by sectors, the percentage of losses due to
wake effects for each sector varies from $\approx 0$\% up to $\approx 1.5$\%. The total
expected annual AEP is equal to $293.274$~GWh, which represents a $4.44$\% reduction due to wake losses.

\begin{table}[!h]
{\scriptsize \caption{\label{Alphaventuslayout}Annual expected AEP considering
wake effects and the actual Alpha Ventus layout.}
\begin{center}
\begin{tabular}{c|cc}
\hline
     &\multicolumn{2}{c}{Original layout}\\
                  \cline{2-3}
& $AEP$  & Wake\\
N. sector & (GWh annual) & ($\%$)\\
\hline
1 & 31.429  &   0.092 \\
2 & 41.497  &   1.507\\
3 & 46.618  &   0 \\
4 & 37.856  &   0 \\
5 & 27.424  &  1.229 \\
6 & 19.713  &  0.072 \\
7 & 10.347  &  0.047 \\
8 & 8.471  &  0.494 \\
9 & 15.038  &   0  \\
10 & 21.703  &   0 \\
11 & 16.631  &   0.931\\
12 & 16.545  &   0.068  \\
\hline
Total & 293.274  &  4.440 \\
\hline
\end{tabular}
\end{center}
}
\end{table}

The proposed methodology is applied to obtain the optimal layout for the 12
turbines on Alpha Ventus. Results are given in Table~\ref{evolucion_opt}, where we present
productions associated with: i) the initial random configuration obtained using the heuristic procedure
${\bfg x}^s$, and ii) the pseudo-global optimum (${\bfg x}^\ast$) obtained throughout the gradient base strategy,
 using ${\bfg x}^s$ as a starting point. According to these results, the following observations are pertinent:
\begin{enumerate}
  \item The optimal expected annual AEP is equal to $304.809$~GWh, reducing the wake effect from
$4.44$\% for the actual layout, to $0.682$\% at the pseudo-global optimum.

\item The maximum loss due to wake effect is reduced from
$1.507$\% for sector 2 and the actual layout, to $0.194$\% for sector 6 at the pseudo-global optimum.

\item The initial solution ${\bfg x}^s$ which allows achieving the pseudo-global optimum, does not necessarily
constitute a good solution per se. Note that losses due to wake
effect for this initial solution are $6.012$\%, higher than those
of the actual configuration.

\item The optimization strategy tends to decrease wake effects for those sector with higher expected AEP.
Note that for sectors 3 and 4, the corresponding wake effects are
null, and for sector 2, they are close to zero.

\item The optimization procedure provides an improvement of expected AEP of
3.758$\%$ with respect to the actual configuration.

\end{enumerate}

\begin{table}
{\scriptsize \caption{\label{evolucion_opt} Annual expected AEP considering
wake effects using the optimization framework proposed in this paper.}
\begin{center}
\begin{tabular}{c|cc|cc}
\hline
         &\multicolumn{2}{|c}{Initial Simulation (${\bfg x}^s$)}&\multicolumn{2}{|c}{Optimal layout (${\bfg x}^\ast$)}\\
                \cline{2-5}
& $AEP_0$& Wake &  $AEP_{opt}$ & Wake\\
N. sector & (GWh annual)& ($\%$) &(GWh annual) & ($\%$)\\
\hline
1 & 29.682  &   0.662  &   31.148   &   0.184\\
2 & 44.522  &   0.522  &   46.035   &   0.029\\
3 & 43.366  &   1.059  &   46.618   &   0\\
4 & 36.054  &   0.587  &   37.856   &   0\\
5 & 29.496  &  0.554  &   31.196   &   0\\
6 & 18.641  &  0.421 &   19.337   &   0.194\\
7 & 9.534  &   0.312  &   10.161   &   0.107\\
8 & 9.447  &  0.176 &   9.958   &   0.010\\
9 & 13.519  &   0.495&    15.038   &   0\\
10 & 20.434  &   0.414 &     21.703   &   0\\
11 & 18.159  &   0.433 &    19.488   &   0\\
12 & 15.595  &   0.377 &    16.269   &   0.158\\
\hline
Total & 288.448  &  6.012    & 304.809&0.682\\
\hline
\end{tabular}
\end{center}
}
\end{table}

\begin{figure}[!h]
\begin{center}
\includegraphics*[width=0.8\textwidth]{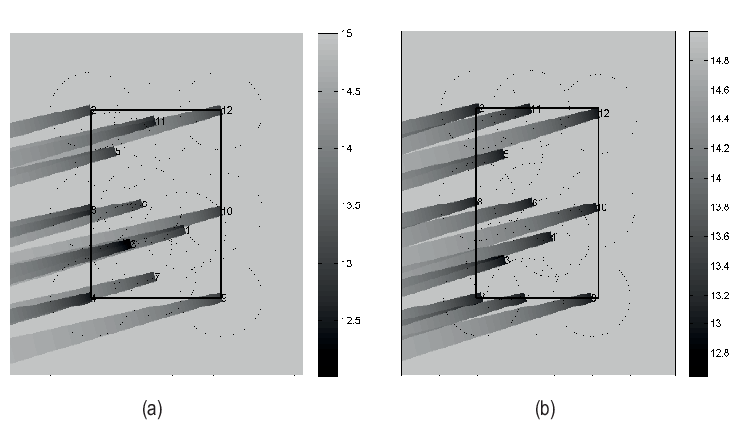}
\caption{\label{estelas_opt_evol} Evolution of the wakes for a
$15$ m/s wind in sector 3 ($75^\circ$NE): a) Initial simulation,
and b) Optimal layout.}
\end{center}
\end{figure}

\begin{figure}[!h]
\begin{center}
\includegraphics*[width=0.8\textwidth]{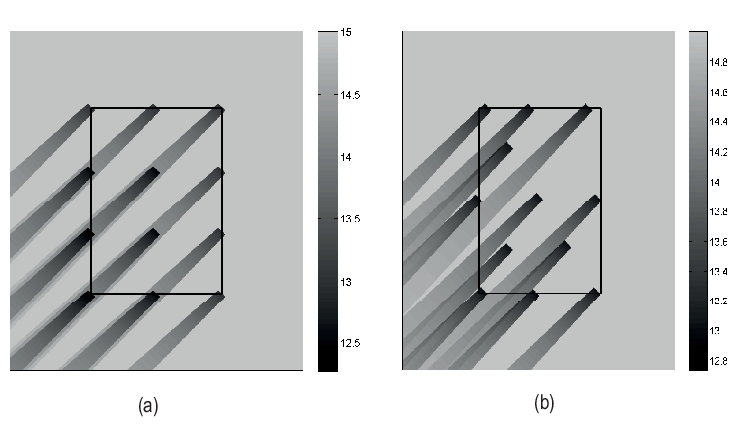}
\caption{\label{estelas_compfinal_1} Evolution of the wakes for a
$15$ m/s wind in sector 2 ($45^\circ$NE): a) Original layout, and
b) Optimal layout.}
\end{center}
\end{figure}

Figure~\ref{estelas_opt_evol} depicts the evolution of wakes for a
specific sector (3, $75^\circ$NE) and wind speed (15 m$/$s), i.e.
the most likely sector. Note that with the initial simulation,
turbines 4, 5, 9, 11 and 12 are under wake effect, however, with
the optimal layout, the original flow speed reaches all turbines.
Figure~\ref{estelas_compfinal_1} shows analogous results to those
of Figure~\ref{estelas_opt_evol} in sector 2 ($45^\circ$NE), for
original and optimal layouts. Note the considerable differences
between the actual configuration and the optimal. 

In economic terms, and assuming a production cost of 0.064
\euro{}/KWh (regardless of the initial investment costs) and a
selling price of 0.15 \euro{}/KWh, the profit increment due to the
reallocation of turbines according to the locations obtained from
the proposed procedure, is equal to $0.9921$M\euro{} per year.
This amount justifies the effort performing the layout
optimization of any off-shore wind farm.

Finally, and in order to study the sensitivity of the proposed
procedure to wind directional data characterization, we have
performed new calculations rotating the original wind rose shown
in Figure~\ref{rosa} with different angles. Results in
Table~\ref{sensibilidad} demonstrate that the optimal layout
always provides better performance with respect to the actual
configuration based on a regular grid. This result is not
surprising as the method tries to accommodate the turbines so that
there are no wake conflicts for each directional sector. Since it
is impossible to avoid wake effects all over the 360$^\circ$, the
method gives priority to those sectors with higher AEPs, but
somehow improves performance throughout the circumference.

\begin{table}
{\scriptsize \caption{\label{sensibilidad} Sensitivity Analysis
with respect to wind directional information for Alpha Ventus case
study.}
\begin{center}
\begin{tabular}{c|cc|cc}
\hline
         &\multicolumn{2}{c}{Original layout}&\multicolumn{2}{|c}{Optimal layout}\\
         \cline{2-5}
 & $AEP$ & Wake & $AEP$ & Wake \\
Rose & (GWh annual)&($\%$)& (GWh annual)& ($\%$) \\
\hline
Rot.0 &293.274 & 4.440& 304.809& 0.682 \\
Rot. 90&293.017 &4.524 & 304.119&0.906\\
Rot. 180& 293.274&4.440 & 304.769&0.695 \\
Rot. 270 &293.017 & 4.524&304.216 &0.875 \\
Random &143.03 & 6.589&148.96 & 2.717\\
 \hline
\end{tabular}
\end{center}
}
\end{table}

\subsection{Efficiency of a Wind Farm: Installed capacity per
km$^2$}
So far, the number of turbines to be allocated has been considered as data. However, this could be
an additional variable to be optimized. To study how the optimal efficiency within a given area changes
 with respect to the number of installed turbines, i.e. installed capacity, we have repeated the optimization process
 using a different number of turbines. Note that the wind farm Alpha Ventus has an actual
installed capacity of 15.625 MW$/$km$^2$. The efficiency achieved
with the optimized layout is 99.25$\%$ versus 95.51$\%$ obtained
with the grid layout.

\begin{figure}[!h]
\begin{center}
\includegraphics*[width=0.8\textwidth]{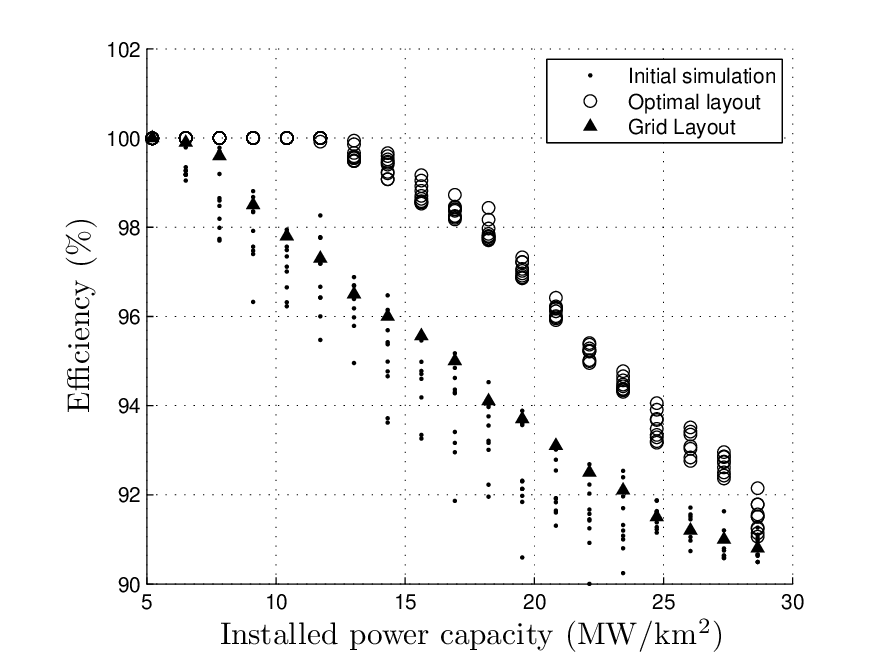}
\caption{\label{eficiencia} Efficiency vs. installed power capacity per
km$^2$.}
\end{center}
\end{figure}

Figure~\ref{eficiencia} shows the efficiencies achieved for
different numbers of turbines in the two stages of the optimization
methodology. It also shows the efficiency achieved with the typical grid layout.
Note that if the installed power capacity per km$^2$ increases, the
efficiency is reduced; in other words, the wake effects become
more important. In addition, we can also observe that the higher the installed power capacity,
 the lower the magnitude of improvement we obtain from the optimization algorithm proposed in this paper.
This effect is due to the wind farm saturation, which reduces the
number of possible turbine locations across the area. Note that
with 23 turbines the problem is unfeasible due to the fact that
there are no layouts which satisfy the problem constraints
(Minimum distances).

Another important reason is that in the N.O. Jensen wake
model, the total velocity deficits inside the wind farm quickly
reach an equilibrium level (see \cite{Katic:86}) and therefore as
the wind farm increases, the layout of the inside turbines start
losing importance, as the wake effects inside the wind farm become
more homogeneous.

\subsection{Saturation area}
\begin{figure}[h!]
\begin{center}
\includegraphics*[width=0.8\textwidth]{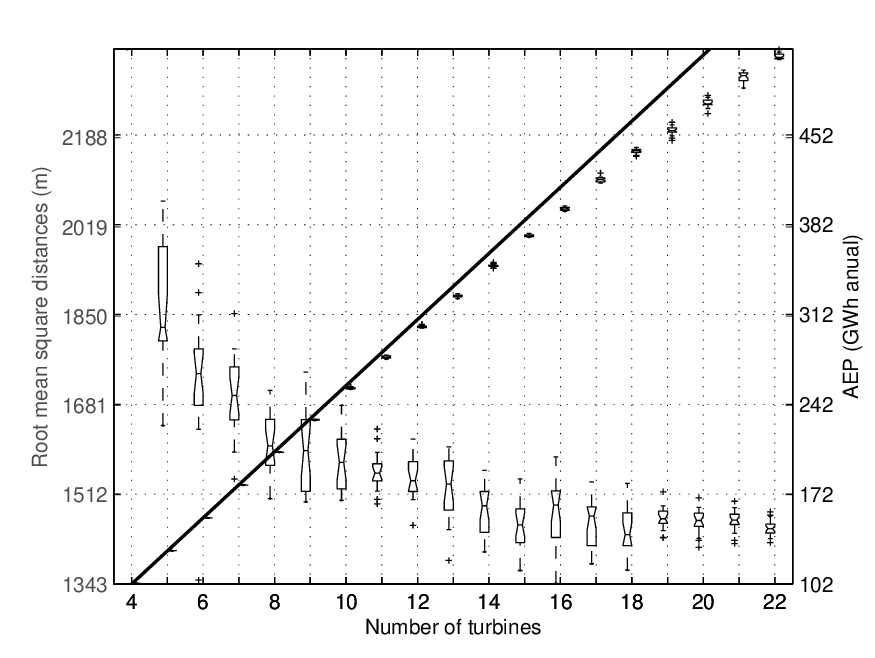}
\caption{\label{boxplot3} Combined box-plots: AEP and root mean square distances.}
\end{center}
\end{figure}

To analyze the saturation effect, Figure~\ref{boxplot3}
illustrates the saturation of the wind farm. This figure shows
box-plots related to: i) the root mean square distances between
the turbine locations, with values converging to a constant around
$1400$ m, and ii) expected annual energy production (AEP), in
which the maximum value is $518$ GWh for 22 turbines. These
box-plots have been generated running 100 cases for each installed
capacity (number of turbines).

The evaluation of the root mean square of distances allows us to
study the difference between the optimal layouts, proving that the
more saturated the area, the solutions are more similar to each
other, resembling the grid layout. Looking at the figure we can
see that for the study area with 19 or 20 turbines, the adjustment
of AEP data ceases to be linear, this is to say, the efficiency
decreases. If an economic study were carried out, approximately
this number of turbines could be the point where the assigned area
reaches its maximum profitability. This could be due to the fact
that the losses induced by wake effect do not compensate the
installation of more turbines. The graph is computed for up to 22
turbines, because with 23 turbines for the study area, there are
no layouts which satisfy the problem constraints.

\subsection{Computational time}
Finally, the performance of the algorithm under
different situations is analyzed by comparing the
estimated complexity of the algorithm with the experimental
computational running times \footnote[1]{Computer characteristic:
Intel Core i5-2400 3.10GHz, 3.24GB RAM running under Windows XP
Professional}.

\begin{figure}[h!]
\begin{center}
\includegraphics*[scale=0.5]{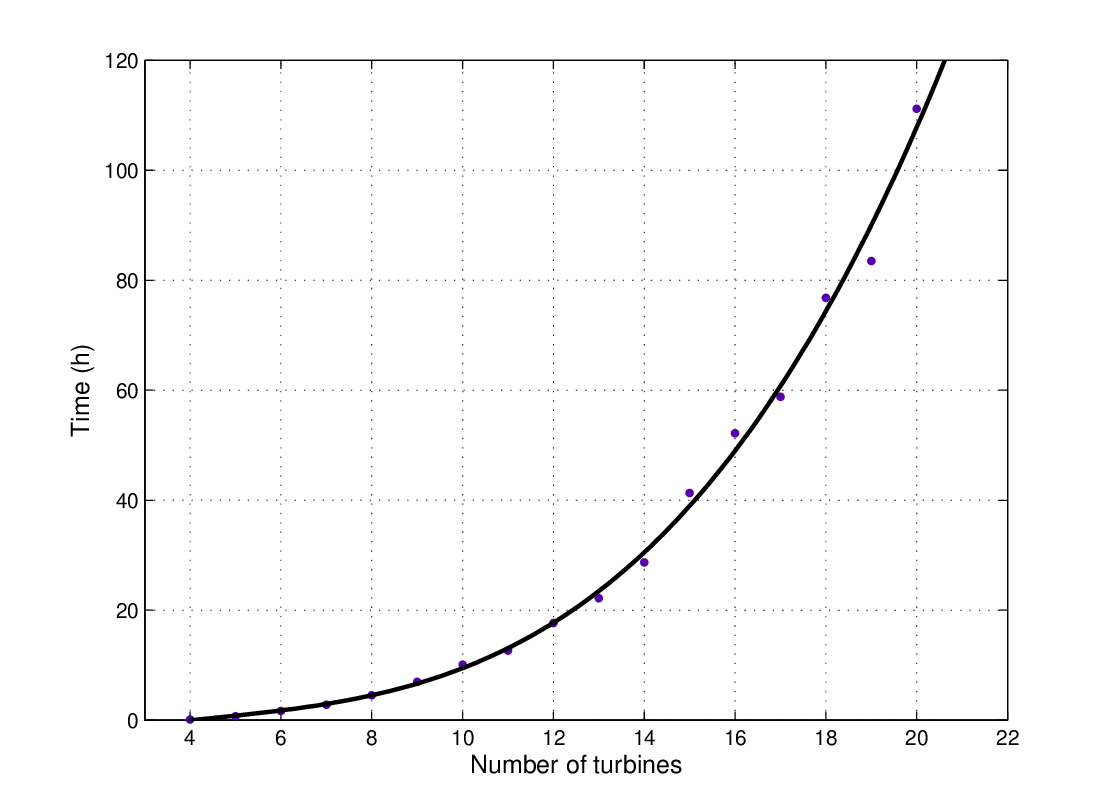}
\caption{\label{tiempo_comp} Running time as a function of
turbines.}
\end{center}
\end{figure}

In Figure \ref{tiempo_comp} the experimental times obtained during
the different test cases are plotted (50 test cases for each
installed capacity).

It is important to note that the above experimental time values
for the different number of turbines were measured during the
development of the program, so not all the simulations began under
the exact same conditions. Therefore, Figure \ref{tiempo_comp}
should be used solely as a reference to know the approximate
running time of the simulation with N turbines.

\section{Conclusions}\label{concl}

This paper proposes a new method to maximize the expected power
production of offshore wind farms by setting the appropriate
layout, i.e. minimizing the wake effects, and indirectly, reducing
the fatigue effects on the turbines, which on turn reduce its
service life.

%

The method uses a
sequential procedure for global optimization consisting of two
steps: an heuristic method to set an initial random layout
configuration, and the use of nonlinear mathematical programming
techniques for local optimization, which use the random layout as
an initial solution.

The performance of the proposed procedure is tested using the
German offshore wind farm Alpha Ventus, yielding an increment of
the expected annual power production of 3.758$\%$ compared to the
actual configuration. A comparison between the installed capacity
per km$^2$ and the efficiency of the wind farm was also carried
out to determine the efficiency of the algorithm when the size of
the wind farm is increased. It was found that when the wind farm
has many turbines and the turbines have little freedom to move
within the wind farm area, the effectiveness of the algorithm is
reduced and the output layout gives similar results as that of any
grid-like layout. On the other hand, for smaller wind farms the
optimization is considerable and the effectiveness of the program
is proven.

The proposed methodology has many advantages: the nonlinear
mathematical programming solvers are numerically robust and
computationally efficient; they can include alternative
constraints or objective functions without altering the flow of
the methodology; the heuristic method used to generate initial
solutions is capable of searching convex subregions; it is easy to
include parallelization features; the final solution holds the
Karush-Kuhn-Tucker optimality conditions; and the methodology does
not requiere reducing the feasible solutions region by gridding
the possible location area.

There is still much work to be done in the field of wind farm
optimization, especially regarding offshore optimization, as it is
very broad and includes wake modeling to properly predict the wind
decay (and therefore the power losses). Other physical aspects
such as foundations or cabling, human aspects as visual effects
and area restrictions, and mathematical topics regarding the
optimization approaches must also still be reviewed in detail.

%
%


\bibliographystyle{elsarticle-num}



\end{document}